\documentstyle[11pt,aussois2003,twoside,epsf]{article}
\def\edcomment#1{\iffalse\marginpar{\raggedright\sl#1\/}\else\relax\fi}
\reversemarginpar
\begin{document}
%
%%---------------------------------------------
%%  please put the title of your talk below :
\title{Cosmic Evolution Impressions on Lensing Maps}
%%----------------------------------------------
%
%%----------------------------------------------
%% your name and affiliation :
\author{Antonio C. C. Guimar\~aes}
\affil{Department of Physics, University of Durham, South Road, Durham DH1 3LE}
%%----------------------------------------------

%%----------------------------------------------
%% if needed, name and affiliation of your co-author :
%\author{My Co-Author}
%\affil{The Name of the Institution, The Full Address of the Institution}
%%-----------------------------------------------
\label{page:first}
\begin{abstract}
I investigate through simulations the redshift dependence of  
several lensing measures for two cosmological models, a flat universe  
with a cosmological constant ($\Lambda$CDM), and an  
open universe (OCDM). 
I argue that quintessence models can be seen  
as an intermediate model between these two models under the weak
gravitational lensing perspective.  
I calculate for the convergence field the angular power spectrum,  
the variance and higher order moments, the probability distribution  
function, and the Minkowski functionals.  
I find that in the redshift range examined (from $z=0.7$ to $z=3$  
various statistics show an increase of the non-Gaussianity of the  
lensing maps for closer sources.   
Lensing surveys of higher redshift are expected to be less non-Gaussian than  
shallow ones because of the non-linear evolution of the density field, 
and the central limit theorem.  
\end{abstract}
\section{Introduction}

A lensed image contains information about the positions of the lens
and source object, and about the lens itself (e.g. its refractive index).
In the cosmological context the mass inhomogeneities in the universe
act as gravitational lenses to source objects far away. 
The dominant cosmological picture today describes a universe that is
expanding and has growing density perturbations. 
The exact characterization and quantification of these phenomena are
essential pursuits of observational Cosmology, and their description
by a theoretical model is a major objective for cosmologists.

Weak gravitational lensing studies can contribute to this enterprise 
because lensing maps depend on the evolution of the large-scale
structure and on the geometrical properties of the universe (the
cosmological expansion). 

In this section I briefly review some of the formalism of cosmic
expansion and structure formation, and point how these two aspects of
cosmic evolution are captured by weak gravitational lensing maps.

\subsection{Cosmological Geometry}

In a Friedmann-Robertson-Walker model, assuming isotropy and
homogeneity, the geometry of the universe can be described by the
metric 
\begin{equation}
d\tau^2 = dt^2 - a^2(t) \left[ dr^2 +f_K^2(r)d\phi^2 \right] \;,
\end{equation}
where $t$, $r$ and $\phi$ are the comoving coordinates, and $f_K$ is 
the curvature-dependent comoving angular distance ($=r$
for a flat universe, $=\sinh r$ for open, and $=\sin r$ for closed). 
The expansion factor $a(t)$ is determined by the equation
\begin{equation}
\dot{a}^2-\frac{8\pi G}{3}\rho a^2 = -k \; ,
\end{equation}
in a universe of curvature $k$ ($=0,-1,1$ in the case
that it is flat, open, or closed, respectively), and of density
$\rho$. In terms of the density parameters $\Omega_\Lambda$,
$\Omega_m$ and $\Omega_r$, for a cosmological constant, matter, and
radiation, respectively, 
\begin{equation}
\rho = \frac{3 H^2_o}{8\pi G} \left( \Omega_\Lambda + \Omega_m a^{-3}
  + \Omega_r a^{-4} \right) \; .
\end{equation}

It is also useful to define the Hubble parameter
\begin{eqnarray}
H & \equiv & \frac{\dot{a}}{a}  \; ,\\
H(a) & = & H_o \left[\Omega_\Lambda (1-a^{-2}) + \Omega_m
  (a^{-3}-a^{-2}) + a^{-2} \right]^{1/2} \; ,
\label{Hubble-param} 
\end{eqnarray}
and to remember the relation $a=1/(z+1)$.

In a quintessence model the cosmological constant is substituted by a
dark energy component, which has negative pressure, i.e. a negative
$w_Q$ in its equation of state $p_Q=w_Q \rho_Q$. 
In the case in which $w_Q$ is constant the Hubble expansion rate is
them
\begin{equation}
H(a)  =  H_o \left[\Omega_\Lambda (a^{-3(w_Q+1)}-a^{-2}) + \Omega_m
  (a^{-3}-a^{-2}) + a^{-2} \right]^{1/2} \; ,
\end{equation}
which reduces to the cosmological constant case [equation
(\ref{Hubble-param})] when $w_Q=-1$.

\subsection{Density Perturbations}

The current standard explanation for the existence of a large-scale structure
in the matter distribution in the universe is that initial
small fluctuations in a remote past evolved (grew) due to gravitational
attraction. That is the gravitational instability scenario.
The origin and characteristics of the initial inhomogeneities is
subject of great 
interest for Cosmology and fundamental Physics, but will not be
discussed here. 

The matter distribution can be described by the density contrast
$\delta=\rho/\bar{\rho}-1$, where $\bar{\rho}$ is the average density,
and its evolution in the linear regime is
given by 
\begin{equation}
\ddot{\delta} + 2 \frac{\dot{a}}{a} \dot{\delta} = 
 4\pi G \bar{\rho} \, \delta
\end{equation}
The growing mode solution for this equation can be factored into 
spatial and temporal functions
\begin{equation}
\delta({\bf x},t) = D(t) \, \delta({\bf x},t_o),
\end{equation}
where the growth function is 
\begin{equation}
D(a) \propto \frac{\dot{a}}{a} \int_0^a \frac{1}{\dot{a}} da \;
\end{equation}
which can be approximated, apart from a normalization factor, by
(Carroll, Press \& Turner 1992; Lahav et al. 1991)
\begin{eqnarray}
D(a) \approx a  \frac{5}{2}  \Omega_m(a) 
\left\{ \Omega_m^{4/7}(a) - \Omega_\Lambda(a) 
\left[ \frac 1{2} + \left( 1+ \frac {\Omega_m(a)}{2} \right)
\left( 1+ \frac {\Omega_{\Lambda}}{70}(a) \right) \right]
\right\}^{-1} \, ,
\label{growth-factor}
\end{eqnarray}
\begin{equation}
\Omega_m(a) = \frac {\Omega_m}{a^3} \left( \frac{H_0}{H} \right)^2 \, ,
\end{equation}
\begin{equation}
\Omega_{\Lambda}(a) = {\Omega_{\Lambda}} \left( \frac{H_0}{H}
\right)^2 \, .
\end{equation}
The growth function must be normalized such that $D(a=1)=1$.  
For an Einstein-de Sitter Universe there is no suppression of
gravitational clustering, and the growth function
reduces to the scale factor, $D(a) = a$.

For a quintessence model (Ma et al. 1999)
$\Omega_m(a)=\Omega_m/[\Omega_m+(1-\Omega_m)a^{-3w_Q}]$,
and the growth factor can be approximated by (apart from a
normalization factor) 
\begin{eqnarray}
D_Q(a) & \approx & (-w_Q)^t a \frac{5}{2}  \Omega_m(a) \\ \nonumber
& \times &
\left\{ \Omega_m^{4/7}(a) -1 + \Omega_m(a) 
+ \left[ 1+\frac{\Omega_m(a)}{2} \right] 
\left[ 1+ \frac{1-\Omega_m(a)}{70} \right]
\right\}^{-1} \; , 
\label{growth-quint}
\end{eqnarray}
\begin{equation}
t=-(0.255+0.305w_Q+\frac{0.0027}{w_Q})[1-\Omega_m(a)]-
(0.366+0.266w_Q-\frac{0.07}{w_Q}) \ln{\Omega_m(a)} \;.
\end{equation}

%********************************************************* FIGURE
\begin{figure}
\leavevmode
\centerline{
\epsfxsize=10.5cm
\epsfbox{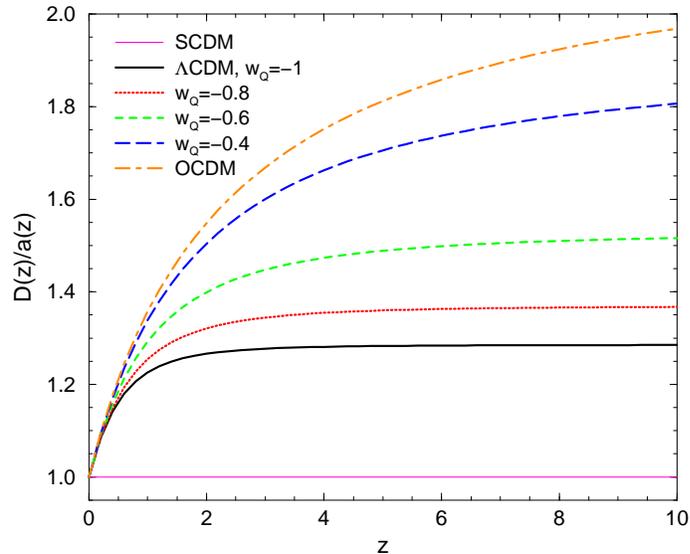}}
%\plotfiddle{growth.eps}{hsf=0.5}
%\plotone{growth.eps}
\vspace{-1.cm}
\caption{Linear growth of factor. SCDM has $\Omega_m=1$, all other
  models have $\Omega_m=0.3$, $w_Q$ is the equation-of-state
  coefficient for a quintessence field model with $\Omega_Q=0.7$.}
\label{growth}
\end{figure}
%********************************************************* FIGURE

I use the expressions (\ref{growth-quint}) and 
(\ref{growth-factor}), properly normalized, to plot the curves in 
Figure \ref{growth} that show the linear growth factor for several
cosmological models. Note that the effect of a quintessence
description for the dark energy component is equivalent to an
intermediate suppression of growth between a flat universe with
cosmological constant and an open universe.

%********************************************************* FIGURE
\begin{figure}
\leavevmode
\centerline{
\epsfxsize=10.5cm
\epsfbox{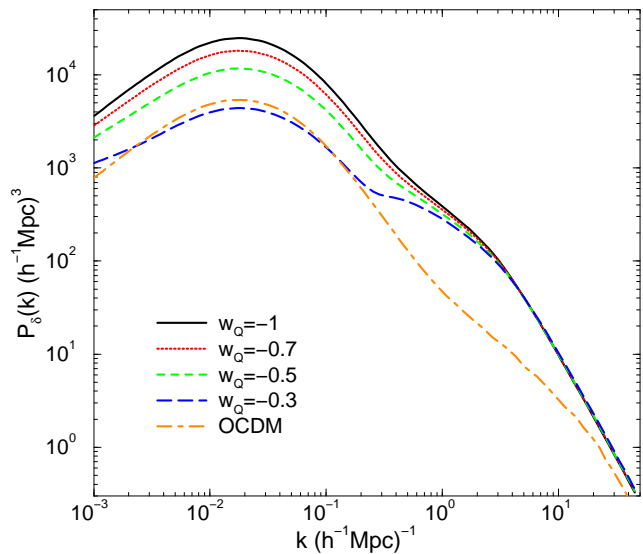}}
\vspace{-1.cm}
\caption{Power spectrum of matter density fluctuations. All models
  (quintessence models and OCDM) were normalized to match the density
  anisotropies observed in the cosmic background radiation (Bun \&
  White 1997 and Ma et al. 1999). The case $w_Q=-1$ is equivalent to a 
  $\Lambda$CDM model.}
\label{Quint_PS}
\end{figure}
%********************************************************* FIGURE

The power spectrum of the matter density perturbation is a powerful
statistical descriptor of the large-scale structure of the universe. 
From linear theory (which is valid at the largest scales) the power 
can be written as
\begin{equation}
P(k,a)=A k^n T^2(k) D^2(a) \;,
\end{equation}
where $A$ is a normalization, $n$ is the initial power spectrum index,
and $T(k)$ is the transfer function, which modifies a initial
power-law spectrum of density perturbations, building in information
about the energy components of the universe.
However, linear theory becomes a bad approximation when 
$\delta({\bf x},t) \gg 1$, which happens at small scales at late
stages of the evolution of the density fields. 
The non-linear evolution causes the power spectrum at large wavenumber
$k$ to be underestimated by linear theory. 
The corrected calculation of the density field evolution in the non-linear
regime requires the use of N-body simulations.
From that one can obtain approximations for the non-linear power
spectrum using prescriptions such the ones developed by Hamilton et
al. (1991), and Scranton \& Dodelson (2000). 
See figure \ref{Quint_PS} for plots of the non-linear power spectrum
of quintessence models with various equation-of-state coefficients, and
of an open universe, with no dark energy.

%\newpage
\subsection{Weak Lensing Dependence on Cosmological Evolution}

The distortion matrix from
which the shear and the convergence fields are defined is explicitly
dependent on the cosmic evolution from the redshift of the source
$z_s$ until today ($z=0$)
\begin{equation}
%\mbox{\boldmath {\theta}}}
{\cal A}_{ij}({\mbox{\boldmath$\theta$}},z_s) = -2 \int_0^{z_s} 
{g(z,z_s) \partial_i \partial_j \Phi({\mbox{\boldmath$\theta$}},z) dz } \; .
\label{evol-distortion}
\end{equation}
This expression integrates the evolution in the geometry of the universe,
which is encoded in the factor $g$, and the evolution of the
gravitational potential $\Phi$.

For the convergence this dependence is represented thought the
comoving distance 
\begin{equation}
w(a) =  
\int _{a}^1 {\displaystyle \frac {da^{\prime}}
{a^{\prime 2} H(a^{\prime})/H_0 }} \; , 
\end{equation}
in the expression
\begin{equation}
\kappa({\mbox{\boldmath$\theta$}},z_s) = \frac{3H_o^2}{2} \Omega_m 
\int_0^{w_s} { \frac{g(w^\prime,w_s)}{a(w^\prime)}
\delta({\mbox{\boldmath$ \theta$}},w^\prime) dw^\prime} \; ,
\label{evol-converg}
\end{equation} 
where
\begin{equation}
g(w^\prime,w)= \frac{f_K(w^\prime) f_K(w-w^\prime)}{f_K(w)} \; .
\end{equation}

%********************************************************* FIGURE
\begin{figure}
\leavevmode
\centerline{
\epsfxsize=10.5cm
\epsfbox{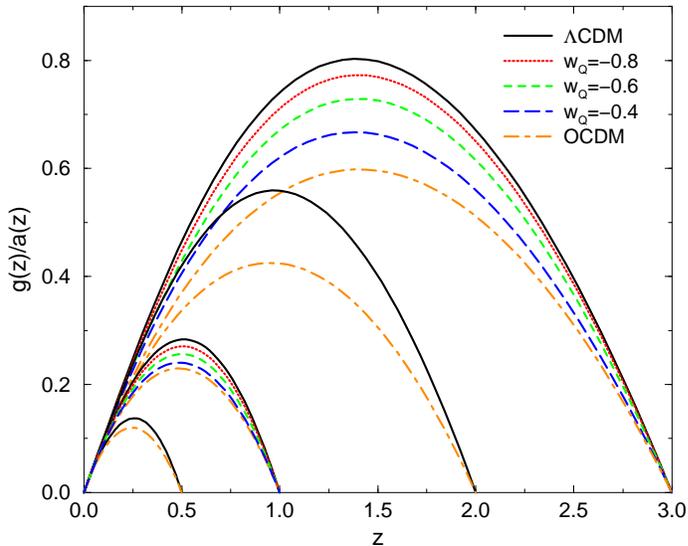}}
\vspace{-1.2cm}
\caption{Geometrical weighting of density perturbations in weak
  lensing. Lines labeled by $w_Q$ represent quintessence models with
  such equation-of-state coefficient.}
\label{weight_z}
\end{figure}
%********************************************************* FIGURE

Figure \ref{weight_z} shows the geometrical weighting of the density
fluctuations for a source at a redshift $z$ given by the factor 
$g/a$ at equation (\ref{evol-converg}).
The maximum contribution for lensing comes from structures with 
redshifts close to half of the source redshift. 
Quintessence models have an intermediate weighting between a
$\Lambda$CDM and a OCDM model, and the farther the source the more
significant is the difference between the models.

%%%%%%%%%%%%%%%%%%%%%%%%%%%%%%%%%%%%%%%%%%%%%%%%%%%%%%%%%%%%%%%%%%%%%%%%%

\section{Lens Evolution}

In this section I use a set of statistics to characterize the evolution of
the matter density field, or more properly, the evolution of lens-planes.
I analyze the projected density contrast field of boxes of 100
$h^{-1}Mpc$ of side that evolve in time. 
The simulations use $128^3$ dark-matter particles (no baryonic
matter) in two models, $\Lambda$CDM and OCDM.

\begin{figure}
\leavevmode
\centerline{
\epsfxsize=11cm
\epsfbox{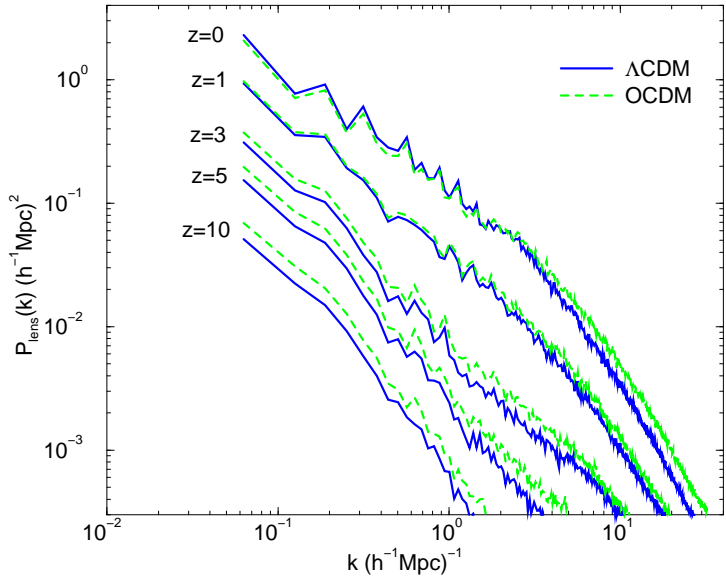}}
\vspace{-1.2cm}
\caption{2D power spectrum of lens-planes.}
\label{power-lens}
\end{figure}

Figure \ref{power-lens} shows the evolution of the two-dimensional
power spectrum of 
the lenses, and illustrates the growth of structure in the matter
density field. 
It is possible to note that small-scale structures (represented by large
$k$) have a stronger increase than large-scale structures (small $k$), as
is expected from non-linear evolution.
The difference between $\Lambda$CDM and OCDM is more notable at large
redshifts, because the simulations were normalized to the cluster
abundance today.

%********************************************************* FIGURE
%********************************************************* FIGURE

\begin{figure}
%\centering 
\leavevmode 
%\gdef\eps@scaling{0.6}
%\includegraphics[width={\eps@scaling\columnwidth}]{stat-lens_z.eps}
%\hspace{-2cm}
%\includegraphics[width={\eps@scaling\columnwidth}]{stat-lens_a.eps}
\centerline{
\epsfxsize=10.5cm
\epsfbox{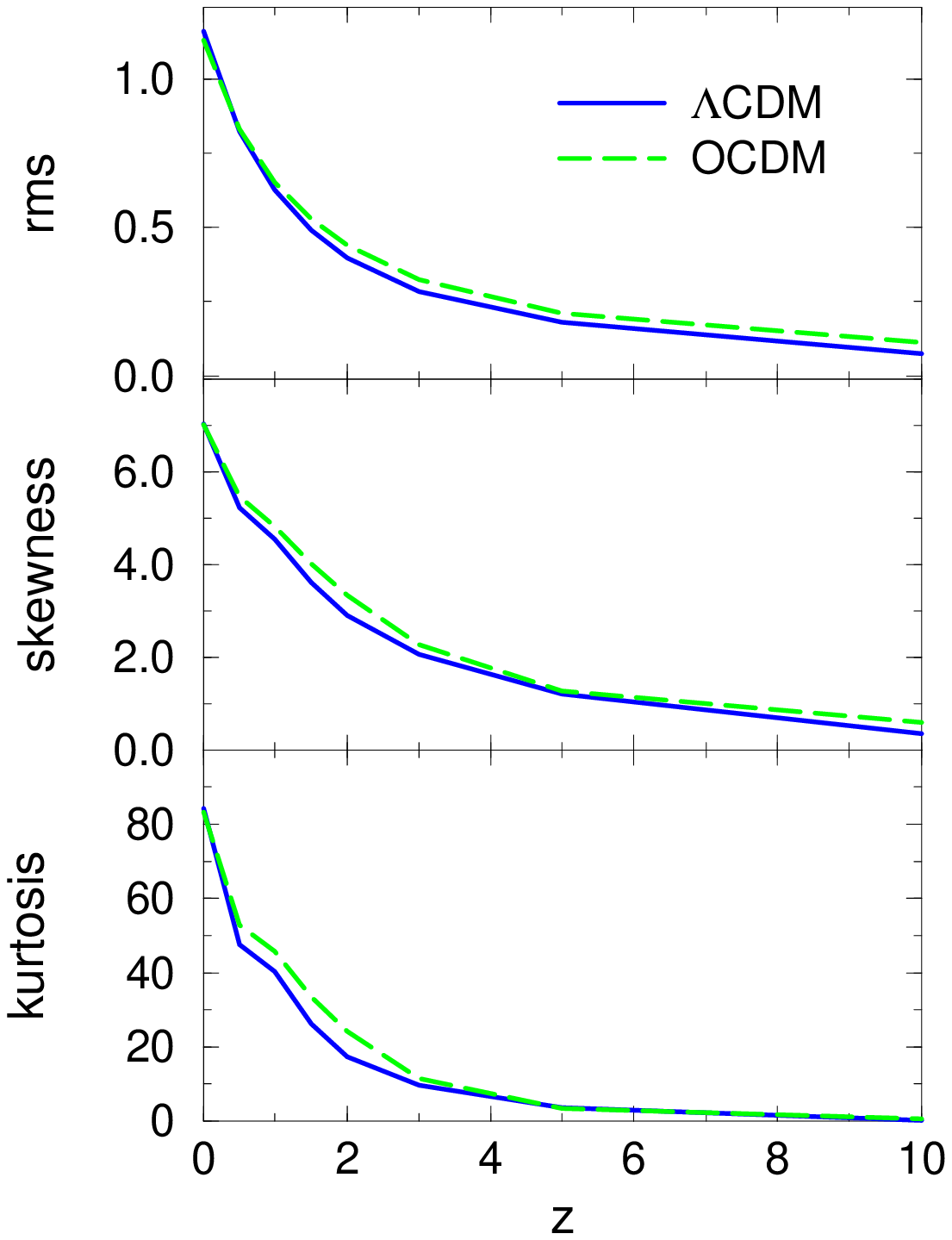}
\hspace{-5cm}
\epsfxsize=10.5cm
\epsfbox{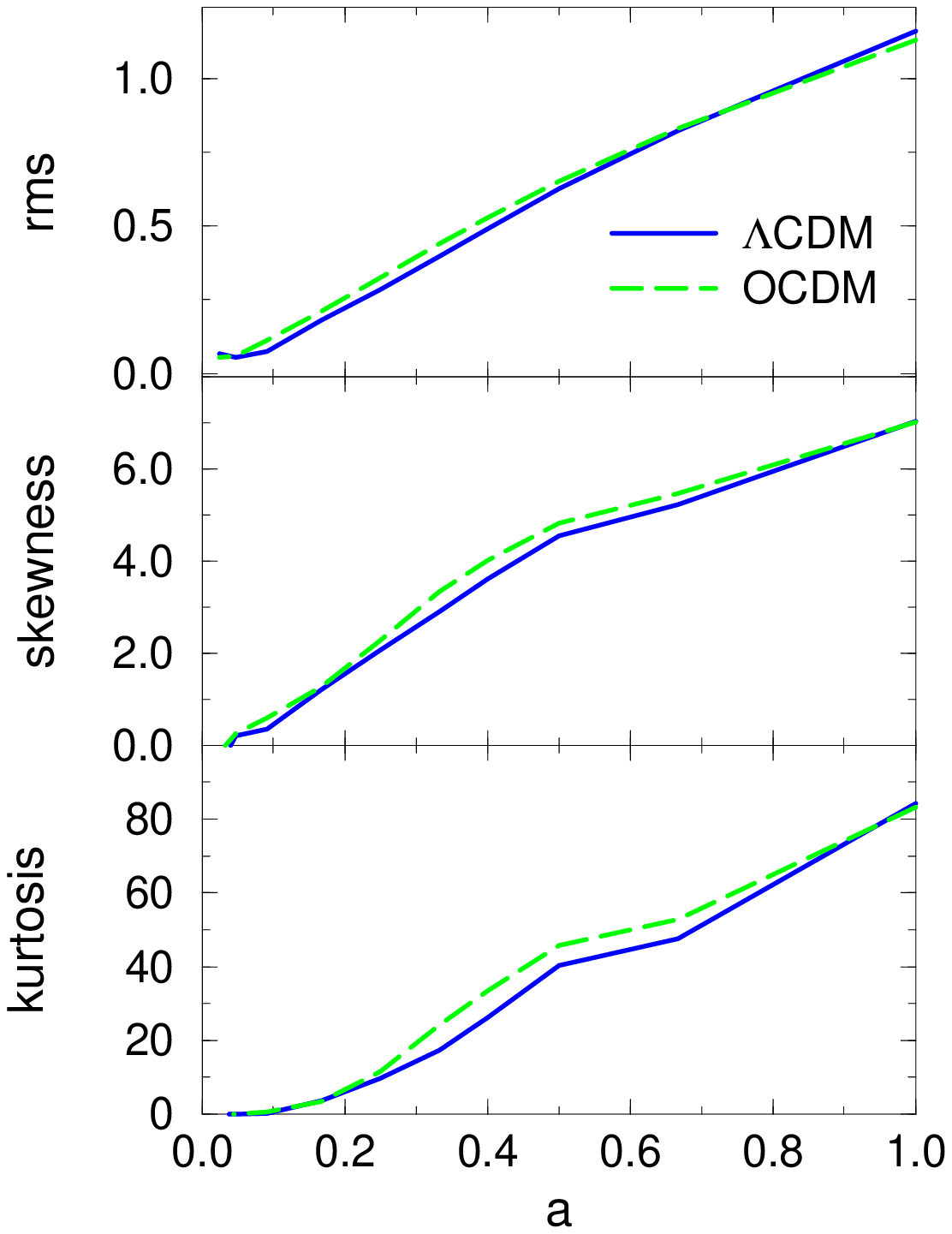}}
\vspace{-1.5cm}
\caption{The rms, skewness, and kurtosis of the projected density
  contrast field as redshift (left) and scale factor (right) functions
 for $\Lambda$CDM (solid lines) and OCDM (dashed 
 lines). Projected boxes have 100 1h$^{-1}$Mpc, and were smoothed at 1
 1h$^{-1}$Mpc scale.}
\label{stat-lens}
\end{figure}

\begin{figure}
\leavevmode
\centerline{
\epsfxsize=10.5cm
\epsfbox{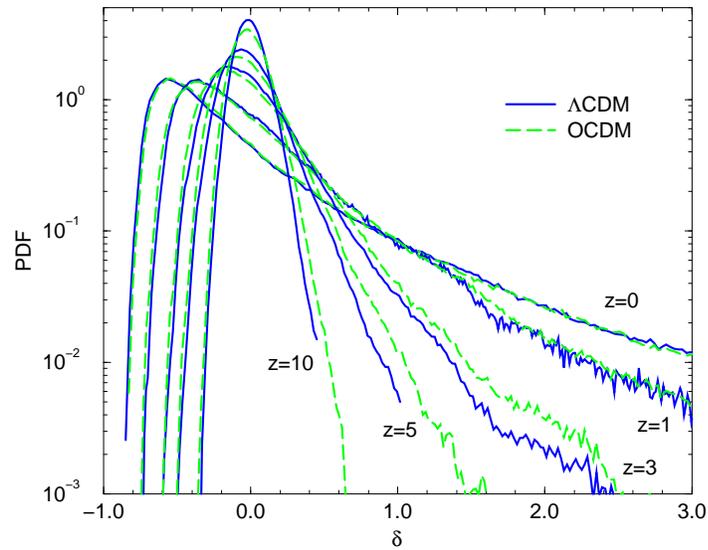}}
\vspace{-1.2cm}
\caption{Probability distribution function for the density contrast of
  lens planes with 1h$^{-1}$Mpc smoothing. } 
\label{pdf-lens}
\end{figure}

The growth of structure in time can be also appreciated by the increase of the
field variance, skewness, and kurtosis - see Figure \ref{stat-lens}. 
These two last quantities quantify the growth of non-Gaussianities in
the density field, which in this case is originally Gaussian, due to
non-linear evolution.
The same information can be obtained from the PDF (figure
\ref{pdf-lens}).

%********************************************************* FIGURE
%********************************************************* FIGURE

%********************************************************* FIGURE
%********************************************************* FIGURE

\begin{figure}
\hspace{-0.5cm}
\epsfxsize=7.0cm
%\centerline{\epsfbox{mink2-lens.eps}}
\epsfbox{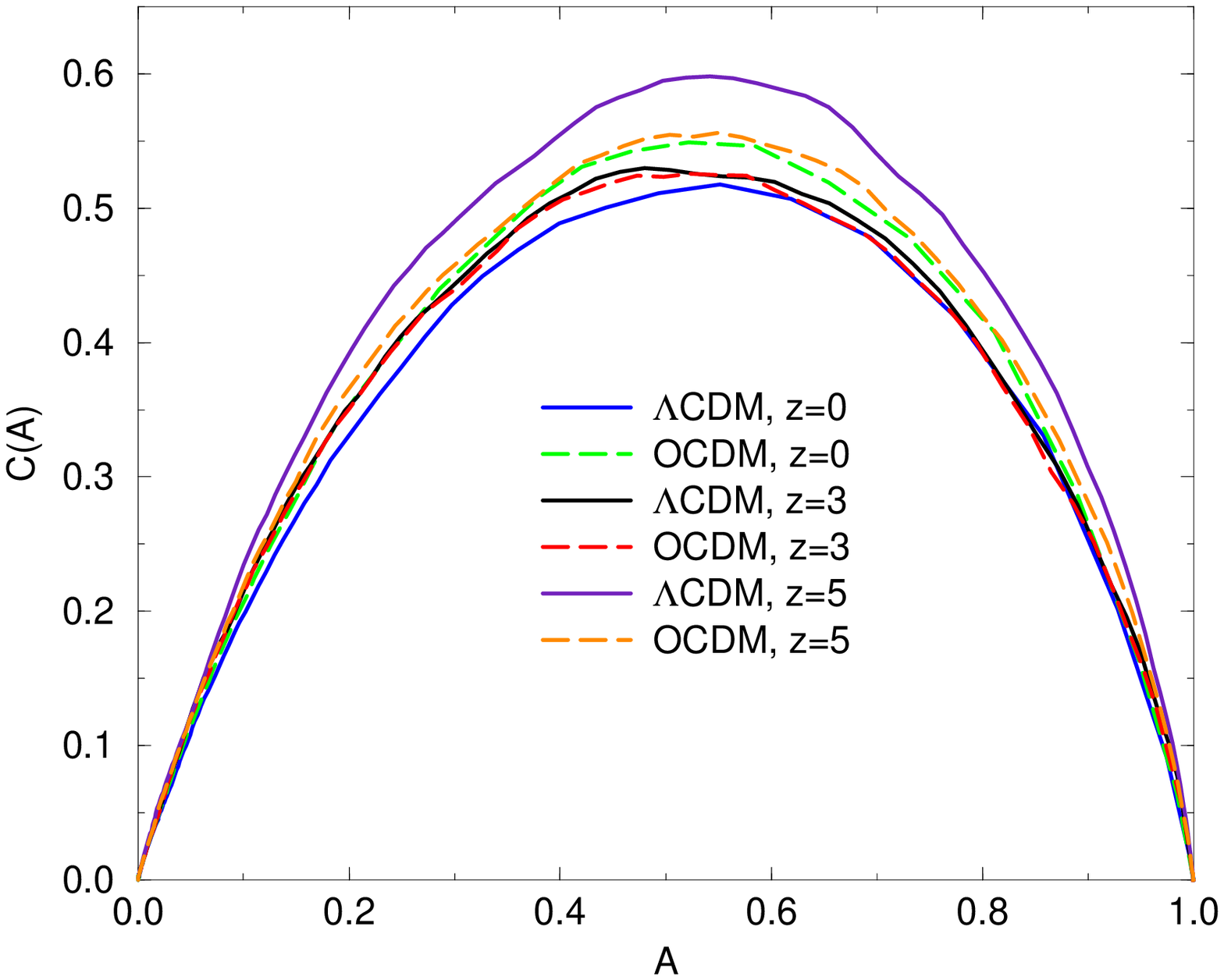}
%\vspace{-1.cm}
\hspace{-0.6cm}
\epsfxsize=7.0cm
%\centerline{\epsfbox{mink3-lens.eps}}
\epsfbox{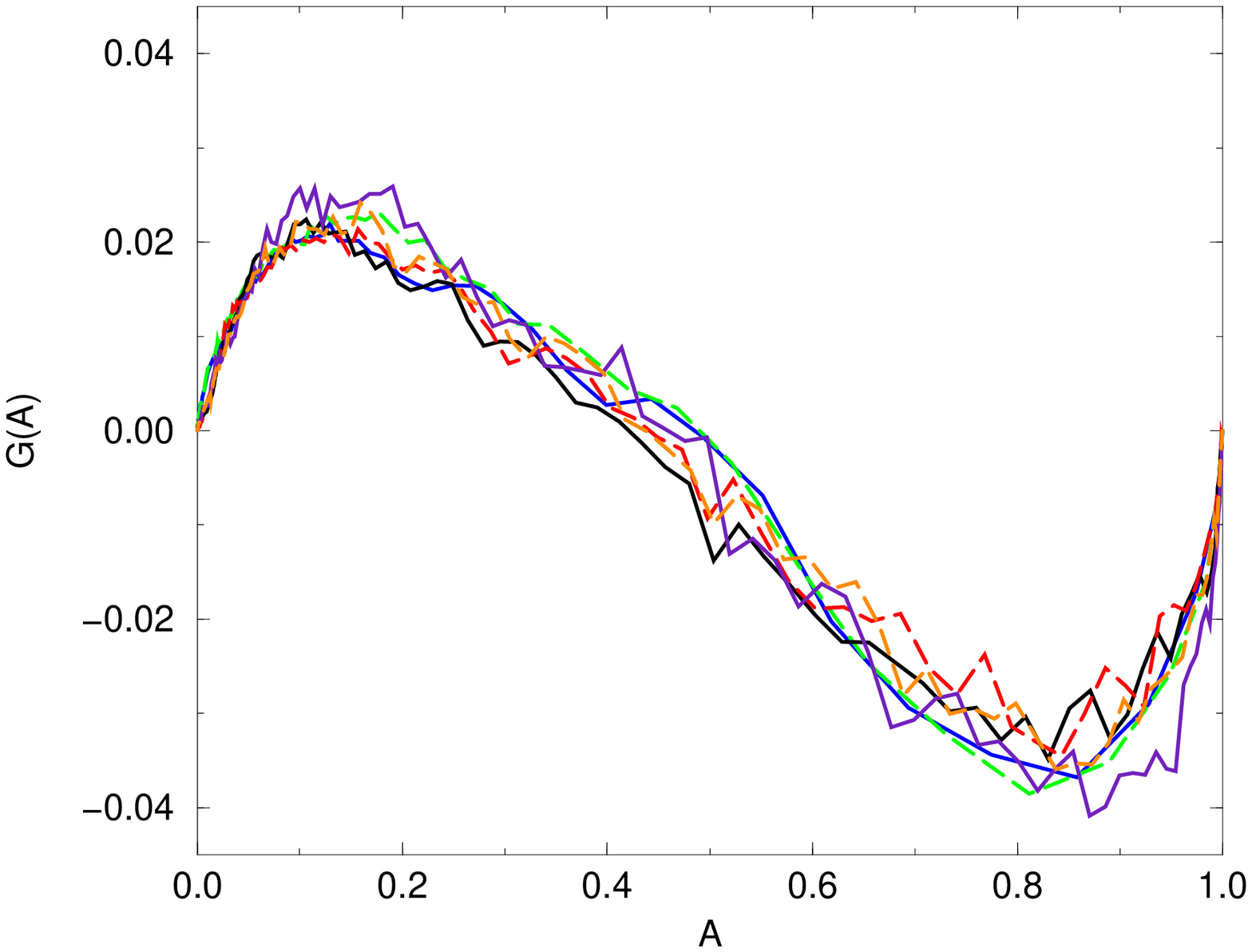}
%\centering 
%\leavevmode 
%\gdef\eps@scaling{0.8}
%\includegraphics[width={\eps@scaling\columnwidth}]{mink2-lens.eps}
%\includegraphics[width={\eps@scaling\columnwidth}]{mink3-lens.eps}
\vspace{-0.7cm}
\caption{Minkowski functionals parameterized by the fractional area $A$
  for the density contrast of lens-planes.}
\label{mink-lens}
\end{figure}

The last aspect of the density field to be analyzed is its
morphology. 
I use the second (contour length) and third (genus) Minkowski
functionals parameterized by the first functional (fractional area) to
characterize the morphology of the lens-planes. 
Figure \ref{mink-lens} shows that up to a redshift z=5 there is
little change in these two functionals.
The third functional is not symmetric in relation to the midpoint
along the axes
of the plot, which again shows the non-Gaussianity of the density
field. 
Both models are almost indistinguishable from one another under this
statistics.
The meaning of these observations is that in both models the
non-trivial morphology, i.e. the morphology distinct from that of a
Gaussian field, emerges at earlier times than z=5, and that in the late
universe the morphology of the LSS as described by the Minkowski 
functionals evolves little if the functionals are parameterized by 
the fractional area, making them independent of the PDF.

%********************************************************* FIGURE
%********************************************************* FIGURE

%%%%%%%%%%%%%%%%%%%%%%%%%%%%%%%%%%%%%%%%%%%%%%%%%%%%%%%%%%%%%%%%%%%%%%%%

\section{Redshift Dependence of Lensing Measures}

%********************************************************* FIGURE
\begin{figure}
\leavevmode
\centerline{
\epsfxsize=14cm
\epsfbox{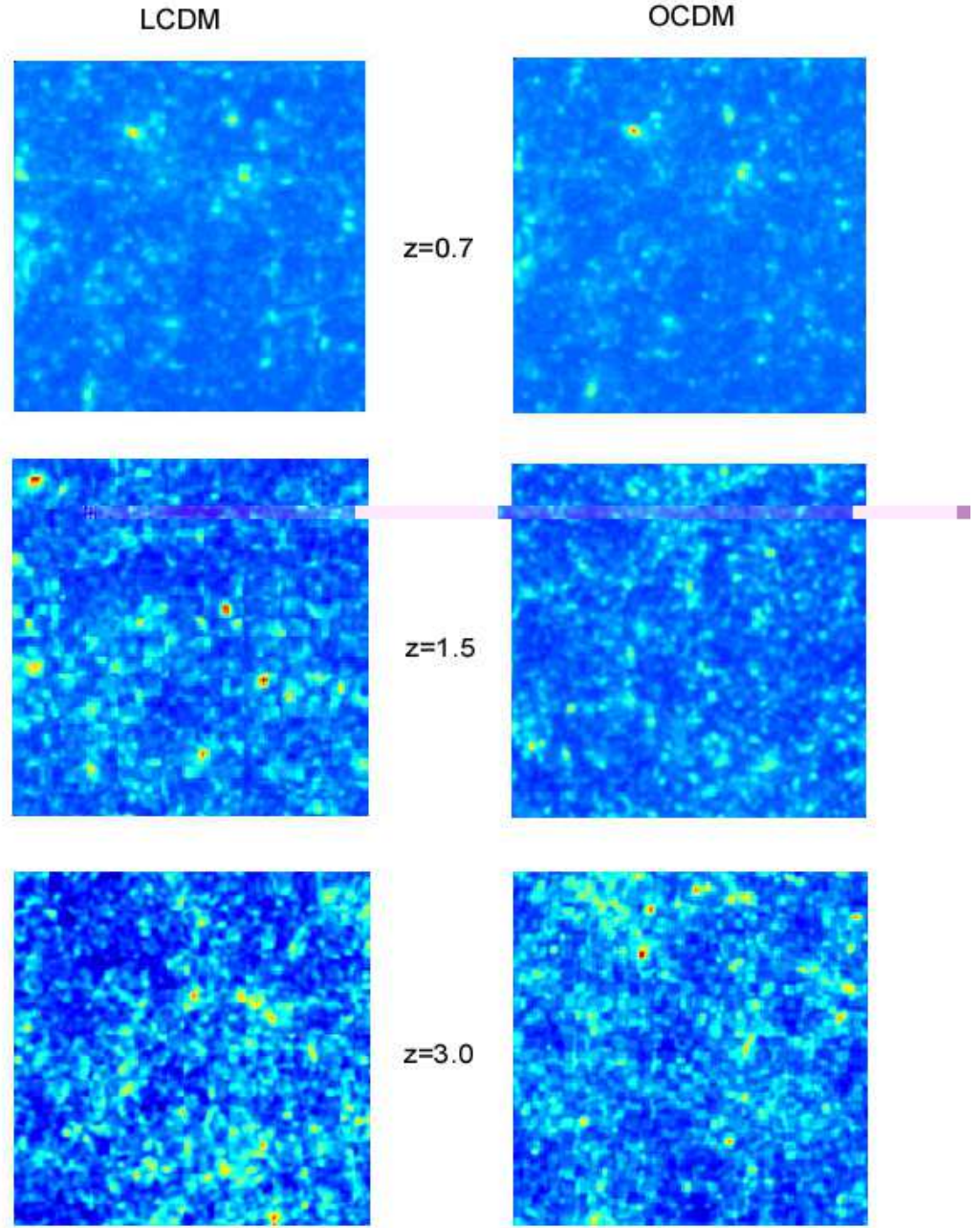}}
\caption{Convergence maps for $\Lambda$CDM (left) and OCDM (right) for
  source plane at redshifts 0.7, 1.5 and 3 (top to bottom). 
  All maps are 1.7$\times$1.7 degrees$^2$, are smoothed with a 1
  arcmin top-hat radius, and use the same color scale (from deep blue for
  the minimum value to red to the maximum value)}
\label{z-kappa}
\end{figure}
%********************************************************* FIGURE

I generated simulated convergence maps using the multiple-plan lens
method (see Guimar\~aes, 2002 for more details) for $\Lambda$CDM and OCDM.
I used the Hydra N-body code (Couchman, Thomas \&
Pearce, 1995) to evolve 128$^3$ particles from a redshift $z=50$ to
$z=0$, simulating the large-scale mass distribution in the two 
models.
For the $\Lambda$CDM ($\Omega_m=0.3,\Omega_\Lambda=0.7$) simulation I
adopt a cluster normalization 
equivalent to $\sigma_8=0.9$, and a simulation box of 141.3
$h^{-1}Mpc$. 
For the OCDM ($\Omega_m=0.3,\Omega_\Lambda=0.7$) simulation I
adopt a cluster normalization 
equivalent to $\sigma_8=0.84$, and a simulation box of 128
$h^{-1}Mpc$.
A Hubble parameter $h=0.7$ was used in both simulations.

Another examined model mimics an universe with the geometry of the 
$\Lambda$CDM model but no structure evolution. In this toy-model 
the structre at any redshift has the same statistical properties 
that today. 

Figure \ref{z-kappa} shows a realization of the convergence map for
both models with a source plane at three redshifts.
The redshift dependence is very clear, but the differentiation between
$\Lambda$CDM and OCDM is less evident.
A quantitative characterization of the models requires the use
of several statistics applied to many realizations of the lensing maps.
I calculate as a function of source redshift the following statistics
for the convergence: the angular power spectrum, the rms (root mean
square), the skewness ($S=\langle{\kappa^3}\rangle /\sigma_\kappa^3$), 
the kurtosis 
$(K=(\langle{\kappa^4}\rangle/\sigma_\kappa^4) - 3$), the probability
distribution function, and the Minkowski functionals.
The results shown in the figures are the result of 20 realizations of
the convergence map for each model, and for each redshift. 
The variance of the measures are shown only for $\Lambda$CDM  
and serves as a reference for maps of same size of other models. 
The sizes of the used maps are (side of square), 4.6$\deg$ for $z=0.7$,
3.5$\deg$ for $z=1$, 2.5$\deg$ for $z=1.5$, 2.2$\deg$ for $z=2$, and 
1.8$\deg$ for $z=3$.

\begin{figure}
\leavevmode
\centerline{
\epsfxsize=12cm
\epsfbox{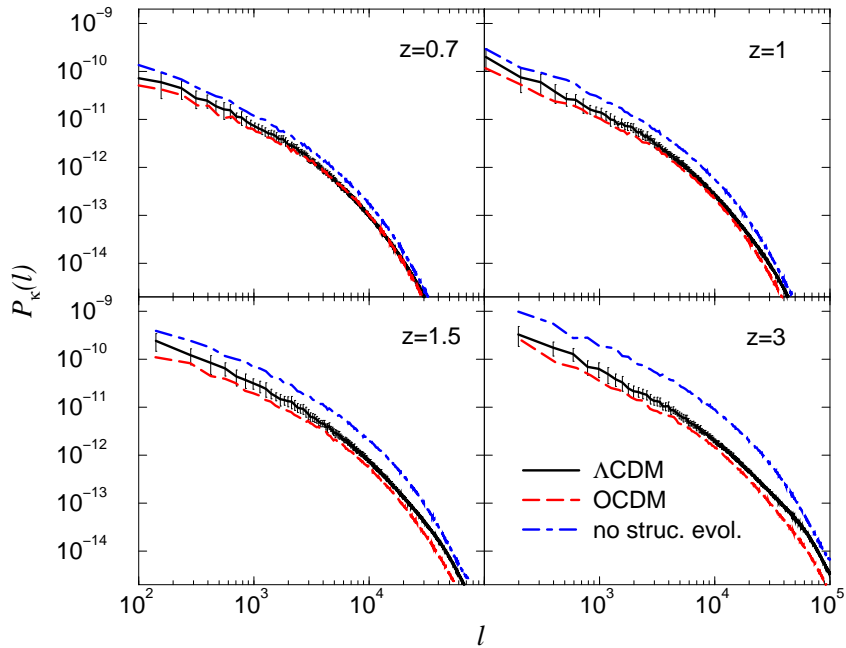}}
\vspace{-1.0cm}
\caption{Angular power spectrum of the convergence field for
  $\Lambda$CDM (left) and OCDM (right) for source plane at redshifts
  3, 2, 1.5, 1, 0.7 (top to bottom curves). No smoothing was used.}
\label{z-power}
\end{figure}

\begin{figure}
\leavevmode
\centerline{
%\vspace{.5cm}
\epsfxsize=11cm
\epsfbox{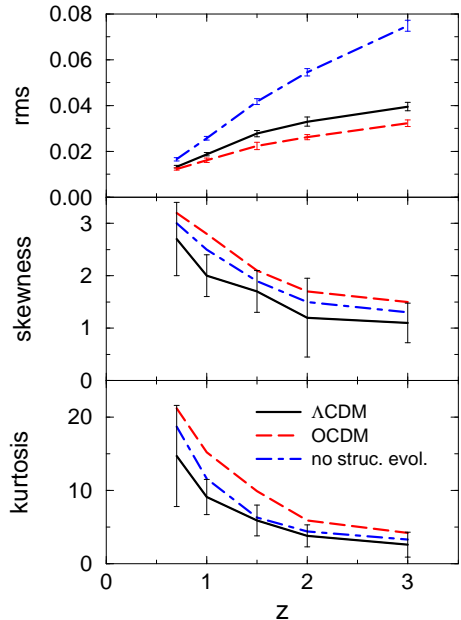}}
\vspace{-1.5cm}
\caption{The rms, skewness, and kurtosis of the convergence field as
  redshift functions for $\Lambda$CDM (solid lines) and OCDM (dashed
 lines). Maps were smoothed at 1 arcmin scale.}
\label{z-statist}
\end{figure}

Figure \ref{z-power} shows the results for the angular power spectrum
of the convergence field. 
There is an overall increase of the spectrum amplitude as the redshift
of the source increases.
This is also reflected in the rms plot (Figure \ref{z-statist}), since
the field variance is proportional to the convergence angular power
spectrum.
The same figure shows a decreasing skewness and kurtosis for larger
redshifts. 
These two statistics can be considered measures of the non-Gaussianity
of the field since for a Gaussian field they are zero.

%********************************************************* FIGURE
%********************************************************* FIGURE

A power law fit to the rms results yield the following redshift
dependence for the shear variance 
\begin{equation}
\langle{\gamma^2}\rangle = \langle{\kappa^2}\rangle \propto z^{\alpha} \; ,
\end{equation}
where $\alpha=1.34\pm 0.08$ for OCDM, and
$\alpha=1.5\pm 0.1$ for $\Lambda$CDM. 
This later result coincides with the value obtained by Jain \& Seljak
(1997), $\alpha=1.52$, but is lower than the value obtained by Barber
(2002), $\alpha=2.07\pm 0.04$.

%********************************************************* FIGURE
%********************************************************* FIGURE

%********************************************************* FIGURE
\begin{figure}
\leavevmode
\centerline{
\epsfxsize=12cm
\epsfbox{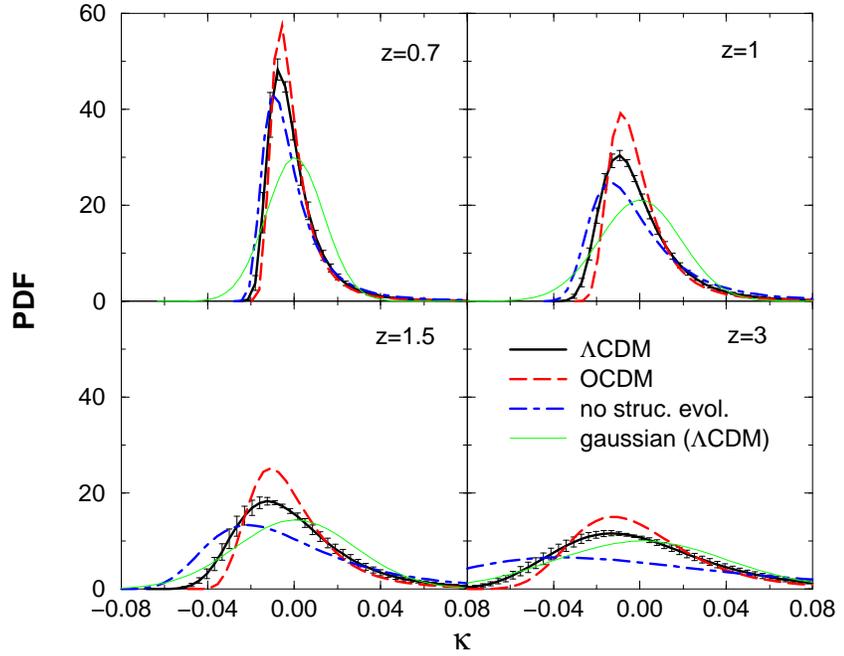}}
\vspace{-1.5cm}
\caption{Probability distribution function of the convergence field
  smoothed at 1 arcmin scale. The dashed lines are the PDF of the
  Gaussianized fields.}
\label{z-pdf}
\end{figure}
%********************************************************* FIGURE

Departures from Gaussianity can also be observed in the probability
distribution function of the lensing maps -- figure \ref{z-pdf}. 
I plot the PDF for the Gaussianized $\Lambda$CDM maps to
show more clearly the non-Gaussianity of the maps. 
The convergence fields were Gaussianized by expanding them in Fourier
modes, and reconstructing the fields with the same mode amplitudes,
but random mode phases.

%********************************************************* FIGURE
\begin{figure}
\hspace{-.3cm}
\leavevmode
\centerline{
\epsfxsize=13cm
\epsfbox{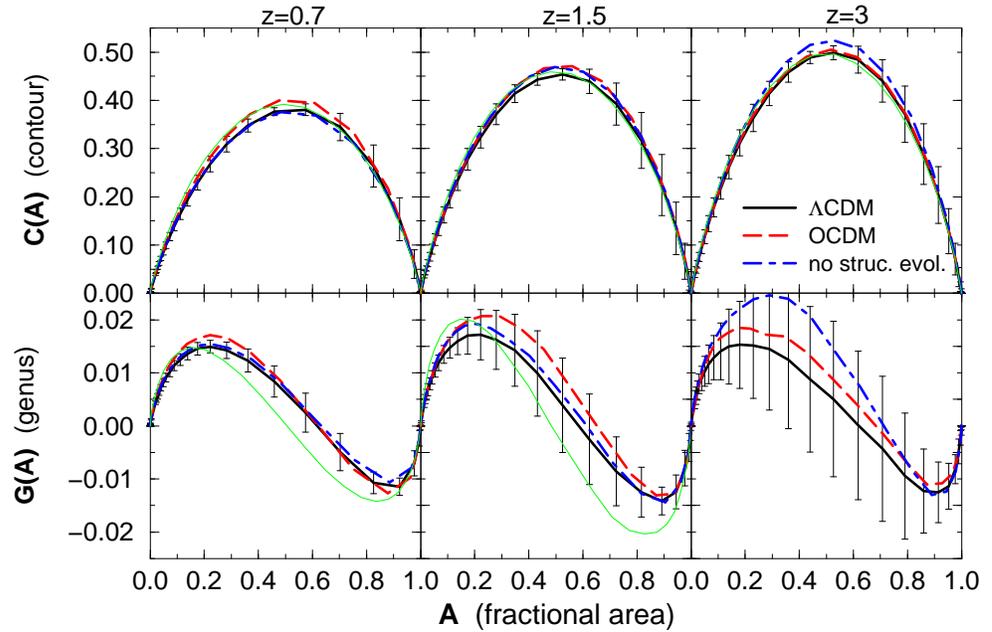}}
\vspace{-0.5cm}
\caption{Minkowski functionals parameterized by the fractional area
  $A$ of the convergence field smoothed at 1 arcmin scale. The thin
  lines are for the Gaussianized maps.}
\label{z-mink}
\end{figure}
%********************************************************* FIGURE

Figure \ref{z-mink} shows the results for the second and third
Minkowski functionals (contour length and topological genus,
respectively), parameterized by the first functional (fractional
area). 
I choose not to present the first Minkowski functional because it
does not carry any extra information in relation to the PDF.
The parameterization by the fractional area $A$ instead of the map
threshold $\nu=\kappa/\sigma_\kappa$ is aimed to further decouple the
information contained in the Minkowski functionals from the PDF.
The results for the Gaussianized fields compared with the original 
field results show that the third functional $G(A)$ puts in evidence the
non-Gaussianity of the maps much more clearly than the second
functional $C(A)$.
Note that the third functional does not show a clear redshift
dependence under the conditions examined.

%%%%%%%%%%%%%%%%%%%%%%%%%%%%%%%%%%%%%%%%%%%%%%%%%%%%%%%%%%%%%%%%%%%%%%
%\newpage
\section{Conclusion}

In this paper I have made predictions for the redshift evolution of
statistical measures of weak lensing maps. 
These predictions show a relatively small difference between OCDM and
$\Lambda$CDM models with adiabatic Gaussian initial conditions, and
will thus also hold in quintessence models in similar conditions.
Given the robustness of the results in redshift dependence within the
class of models tested, I propose the study of the redshift dependence
of weak gravitational lensing as a test for the stability of our
current paradigm for the formation of galactic structure by
gravitational instability in a theory with a primordial scale
invariant spectrum of Gaussian perturbations.

This work concentrates on the convergence field, which can be seen as
a weighted projection of the density contrast field until the
source.
The characteristics of the lensing map will them be dependent not only on
the density field, but on how it is weighted. 
I show how these two components relate to the cosmic evolution,
and what are the resulting lensing properties for two cosmological
models.

The differentiation of the lens-planes is small for the two models
considered when the spectrum of matter fluctuations is normalized to
match the abundance of rich clusters today.
However the lensing geometrical factor, the term $g/a$, is clearly
distinct between the two models, allowing the clear differentiation of
the models by a power spectrum analysis, or by one-point statistics.
It is to be expected, therefore, that for quintessence models these
statistics will assume values intermediate between what was obtained
for $\Lambda$CDM and OCDM, because the geometrical factor is
intermediate. 

It would be very useful if the most popular N-body codes, for example
Hydra (which was used here), could incorporate quintessence
models. 
That would allow a direct calculation of lensing measures in this
model.

The morphological analysis of the maps indicates some evolution of the
second Minkowski functional (equal) for both models, and no
distinguishable evolution of the third functional (genus) in the
redshift range explored. 
The third functional is a better indicator of the departure from
Gaussianity than the second functional. 
Because the third Minkowski functional depends weakly
on the redshift of the source it is a specially recommended statistic
to evaluate Gaussianity in surveys that have a poor redshift
determination of the sources.

%\acknowledgements

%
\label{page:last}
\end{document}